\documentclass[final]{main}

\usepackage{times}
\usepackage{epsfig}
\usepackage{graphicx}
\usepackage{amsmath}
\usepackage{amssymb}
\usepackage{booktabs}
\usepackage{pifont}
\usepackage{multirow}
\usepackage[numbers]{natbib}
\usepackage{makecell}
\usepackage{xcolor}
\usepackage{ifsym}
\usepackage{etoolbox}

\usepackage{soul}
\usepackage{amsmath}
\usepackage{amssymb}
\usepackage{microtype}
\usepackage{wrapfig}
\usepackage{fancyhdr}

\def\figurePath{figures/}
\def\myfigure#1#2{\begin{figure}[htb]\centering\includegraphics*[width = \linewidth]{\figurePath#1}\centering\caption{#2}\label{fig:#1}\end{figure}}

\def\mycfigure#1#2{\begin{figure*}[t]\centering\includegraphics*[clip, width = \linewidth]{\figurePath#1}\centering\caption{#2}\label{fig:#1}\end{figure*}}

\renewcommand{\eg}{e.\,g., }
\renewcommand{\ie}{i.\,e., }
\renewcommand{\etal}{et~al.\ }

\newcommand{\refSec}[1]{Sec.~\ref{sec:#1}}
\newcommand{\refFig}[1]{Fig.~\ref{fig:#1}}

\newcommand{\refTbl}[1]{Tbl.~\ref{tbl:#1}}

\soulregister\ref7
\soulregister\cite7
\soulregister\refFig7
\soulregister\cite7
\soulregister\ref7
\soulregister\pageref7
\soulregister\shortcite7
\soulregister\eg0
\soulregister\ie0
\soulregister\etal0

\DeclareGraphicsExtensions{.png,.jpg,.pdf,.ai,.psd}
\newcommand{\mysection}[2]{\section{#1}\label{sec:#2}}
\newcommand{\mysubsection}[2]{\subsection{#1}\label{sec:#2}}
\newcommand{\mysubsubsection}[2]{\subsubsection{#1}\label{sec:#2}}

\definecolor{colorA}{HTML}{777777}
\definecolor{colorB}{HTML}{4285f4}
\definecolor{colorC}{HTML}{ea4335}
\definecolor{colorD}{HTML}{fbbc04}
\definecolor{colorE}{HTML}{34a853}
\definecolor{colorF}{HTML}{ff6d01}
\definecolor{colorG}{HTML}{46bdc6}
\definecolor{colorH}{HTML}{000000}

\newcommand{\colorIcon}[2]{\textcolor{color#1}{\csname icon#2\endcsname}}
\newcommand{\colorIconAndName}[2]{\colorIcon{#1}{#2} \text{#2}}
\newcommand{\nameInColor}[1]{\textcolor{color#1}{\text{#1}}}
\newcommand{\nameAndIcon}[1]{``#1'' ({\csname icon#1\endcsname})}
\newcommand{\nameAndIconA}[1]{(``#1'', {\csname icon#1\endcsname})}

\usepackage[pagebackref=true,breaklinks=true,colorlinks,bookmarks=false]{hyperref}

\newcommand{\myparagraph}[1]{\vspace{0.15cm} \noindent\textbf{#1}\ \ }

\newcommand{\method}[1]{\texttt{#1}}

\newcommand{\cmark}{\checkmark}
\newcommand{\xmark}{\scalebox{0.95}{\ding{53}}}

\colorlet{colorDirect}{colorA}
\colorlet{colorBM3D}{colorB}
\colorlet{colorFFDNet}{colorC}
\colorlet{colorDBGAN}{colorD}
\colorlet{colorSRNDB}{colorE}
\colorlet{colorLSD}{colorF}
\colorlet{colorHeide}{colorG}
\colorlet{colorOur}{colorH}

\newcommand{\methodCell}[2]{\multirow{#1}{*}{#2}}
\newcommand{\nacell}{\multicolumn1r{-----}}

\newcommand{\mymath}[2]{\newcommand{#1}{\TextOrMath{$#2$\xspace}{#2}}}

\mymath{\network}{G}

\newcommand{\task}[1]{\textsc{#1}}

\newcommand{\clean}{\textsc{Clean}\xspace}
\newcommand{\distorted}{\textsc{Distorted}\xspace}
\newcommand{\cleanToDistorted}{\textsc{Clean}$\rightarrow$\textsc{Distorted}\xspace}
\newcommand{\distortedToClean}{\textsc{Distorted}$\rightarrow$\textsc{Clean}\xspace}

\begin{document}

\title{HDR Denoising and Deblurring by Learning Spatio-temporal Distortion Models}

\author{%
Uğur Çoğalan$^1$\ \ \ \ 
Mojtaba Bemana$^1$\ \ \ \ 
Karol Myszkowski$^1$\ \ \ \ 
Hans-Peter Seidel$^1$\ \ \ \ 
Tobias Ritschel$^2$
\\
\\
$^1$MPI Informatik\qquad
$^2$University College London
}

\maketitle

\begin{abstract}
We seek to reconstruct sharp and noise-free high-dynamic range (HDR) video from a dual-exposure sensor that records different low-dynamic range (LDR) information in different pixel columns:
Odd columns provide low-exposure, sharp, but noisy information; even columns complement this with less noisy, high-exposure, but motion-blurred data.
Previous LDR work learns to deblur and denoise (\distortedToClean) supervised by pairs of \clean and \distorted images.
Regrettably, capturing \distorted sensor readings is time-consuming; as well, there is a lack of \clean HDR videos.
We suggest a method to overcome those two limitations.
First, we learn a different function instead: \cleanToDistorted, which generates samples containing correlated pixel noise, and row and column noise, as well as motion blur from a low number of \clean sensor readings.
Second, as there is not enough \clean HDR video available, we devise a method to learn from LDR video instead.
Our approach compares favorably to several strong baselines, and can boost existing methods when they are re-trained on our data.
Combined with spatial and temporal super-resolution, it enables applications such as re-lighting.
\end{abstract}

\mysection{Introduction}{Introduction}
Common cameras only capture a limited range of luminance values (LDR), while many display and editing tasks would greatly benefit from capturing a higher range of luminance values (HDR) \cite{reinhard2010high}. 
Modern sensors, such as some CMOSIS CMV and Sony IMX sensors, 
allow one to configure different levels of exposure for different spatial patterns \cite{cmosis2020,sony2020}.
This allows HDR by spatial interleaving of different exposures across the sensor.
The challenge is to combine different exposures into a coherent natural image (\refFig{Teaser}).

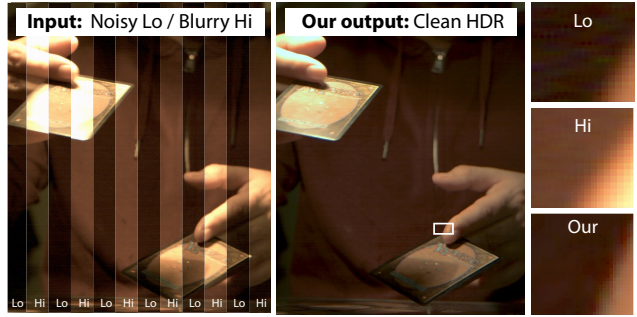
\myfigure{Teaser}{Our method maps low-exposure LDR data with noise and high-exposure LDR data with blur into a clean HDR image.}

Let us consider, without loss of generality, a case where every even row column is captured with a low exposure and every odd row column with a high exposure.
This leads to three specific distortions:
First, \emph{pixel noise} inside the image does not follow a single model anymore, but is now strongly correlated with the column.
Different exposures lead to different noise, one of the reasons why different exposures are being used in the first place: the low exposures have high noise, but are not clamped, while the high exposures have less noise but suffer from clamping.
Second, such cameras suffer from increased levels of \emph{row/column noise}, so orthogonal to the exposure layout, entire rows/columns of pixels change coherently, and differently for different exposures.
Third, and most different from other sensors, the different exposure level also leads to different forms of \emph{motion blur} (MB).
Not only does MB lead to spatially varying blur, but this blur rapidly alternates between odd and even columns.
Low exposures have low MB, while high exposures suffer from strong MB.
In summary, these distortions do not follow any common noise or motion blur model, and hence no method making such assumptions is applicable to HDR from dual exposure.

Removing image distortions (deblurring and denoising) is  now typically solved \cite{zhang2018ffdnet,mao2016image,nah2017deep,tao2018scale} by learning a deep neural network (NN) such as a convolutional neural network (CNN) to implement \distortedToClean.
In our case, this is difficult, as capturing \distorted sensor readings is time-consuming, and there is also a lack of \clean HDR videos.
We suggest a method to overcome both limitations.

Addressing the first, we learn a different function instead: \cleanToDistorted, which generates samples containing correlated pixel noise, row and column noise, as well as motion blur from  \clean sensor readings.
Previous work has made simplifying assumptions, such as Gaussian or Poisson noise, none of which apply to our problem.
We suggest a non-parametric noise model that is  expressive, yet can be trained on a low number of \clean-\distorted pairs.

Second, as there are not enough \clean samples which require HDR video, we supervise from LDR video instead.
Unfortunately, this LDR video does not have the same type of MB as found in HDR sensor readings.
Hence, we use high-speed LDR video to simulate column-alternating MB.

Our evaluation shows that this synthetic training data drives our network, resulting in state-of-the-art HDR images, but can also boost existing methods, including vanilla non-learned denoisers like BM3D, when re-tuned.
Applications span different exposure ratios, where we show re-lighting in a VR/AR context as a typical HDR application.

\mysection{Previous work}{PreviousWork}
In this section, we discuss previous approaches to single (\refSec{SingleImage}), multiple (\refSec{MultiImage}), and in particular HDR (\refSec{HDRImages}) image denoising and deblurring.

\mysubsection{Single-image denoising and deblurring}{SingleImage}

\myparagraph{Noise modeling}
Classic solutions involve fitting Gaussian and Poisson \cite{healey1994radiometric,liu2007automatic} or more involved  \cite{ploetz2017benchmarking} distributions, sometimes under extreme conditions \cite{chen2018learning}, to many pairs of \clean and \distorted images.
While parametric noise models routinely are used as mathematically tractable priors, we use more expressive non-parametric models, as all we need is to generate distorted training data.

\myparagraph{Denoising}
Denoising has traditionally been performed directly on noisy images using state-of-the-art algorithms such as BM3D \cite{dabov2007image}, non-local means \cite{buades2005non}, and Nuclear Norms \cite{gu2014weighted}.
Most deep denoisers \cite{chen2018learning,zhang2018ffdnet,zhang2017gaussian,burger2012image,mao2016image,chen2018imagea,guo2019convolutional,lefkimmiatis2018universal,jia2019focnet} are simply trained on pairs of noisy and clean images, while some work is trained without pairs \cite{ulyanov2018deep,lehtinen2018noise2noise,krull2019noise2void,laine2019high,krull2019probabilistic,batson2019noise2self,quan2020self2self,moran2020noisier2noise,xu2019noisy}, using GANs \cite{chen2018image} or self-supervision \cite{wu2020unpaired}.
The usefulness of neural networks in denoising for real sensors has been disputed \cite{ploetz2017benchmarking,chen2018learning}.

\myparagraph{Blur modeling}
Video obtained with a high-speed camera \cite{su2017deep,nah2017deep,nah2019ntire} or beam splitters \cite{zhong2020efficient} enables motion blur synthesis for the purpose of generating training data using gyroscope-acquired \cite{mustaniemi2020lsd$_2$} or random \cite{mildenhall2018burst} motion.


\myparagraph{Deblurring}
Non-blind deconvolution methods \cite{zoran2011fromlearning,schuler2013amachine,libin2014good,schmidt2013discriminative,xu2014deep,cho2011handling,whyte2010non} restore sharp images given the blur kernel.
Blind deconvolution methods attempt to derive the kernel based on various priors on either the sharp latent image or the blur kernel \cite{fergus2006removing,levin2009understanding,xu2010two-phase,michaeli2014blind,gong2017frommotion,sun2015learning,chakrabarti2016aneural}.
Explicit kernel derivation can be avoided in end-to-end training, where  the sharp image is derived directly \cite{nah2017deep,tao2018scale}, by self-supervision \cite{liu2020self} or adversarial training \cite{kupyn2018deblurgan,kupyn2019deblurgan}.
Video deblurring additionally capitalizes on inter-frame relationships, while assuring temporal coherence of the result \cite{kim2015generalized,kim2017online,zhou2019spatio,zhong2020efficient,su2017deep}.
Deblurring can be combined either with spatial \cite{zhang2019gated} or temporal \cite{purohit2019bringing,jin2018learning,jin2019learning} super-resolution, as done in our approach.
The presence of noise, clamping and multiple exposure as in our condition adds a further challenge.
Methods such as \citet{pan2020physics} model general distortions using CycleGAN \cite{zhu2017unpaired}, but have not been demonstrated to perform denoising.

\mysubsection{Multi-image denoising and deblurring}{MultiImage}
A number of solutions have been proposed to capture multiple images of the same content to provide more information for ill-posed deblurring and denoising. 


\myparagraph{Fixed-exposure burst photography}
Burst photography combines a handful of low-exposure frames into a high-quality LDR result using efficient hand-crafted solutions deployed in cellphones \cite{liu2014fast,hasinoff2016burst,liba2019handheld,liba2019handheld} or based on learning of recurrent architectures \cite{wieschollek2017learning}, or unordered sets \cite{aittala2018burst}, or per-pixel filter kernels \cite{mildenhall2018burst}.
The problem of read noise that accumulates from each contributing frame can be avoided in quanta burst photography that employs binary single-photon cameras to  capture high-speed sequences \cite{ma2020quanta}.

\myparagraph{Low/high exposure image pairs}
Short-exposure images are sharp but noisy, while long-exposure images are blurry but free of noise.
Such \emph{exposure pairs} have been used for non-uniform kernel deblurring \cite{yuan2007image,whyte2010non}.
Along a similar line, \citet{mustaniemi2020lsd$_2$} and \citet{chang2020low} jointly learn how to denoise and deblur exposure pairs supervised by synthetic training data.
Different from our goal, they produce LDR output, while we aim for HDR.

\mysubsection{HDR images and video}{HDRImages}
HDR means covering a large range of luminance via software expansion, multiple exposure, or special sensors.

\myparagraph{Dynamic range expansion}
LDR can be expanded to HDR in software.
Although immense progress has been made based on CNNs \cite{marnerides2018expandnet,endo2017deep,eilertsen2017hdr}, results do not yet match the quality of multi-exposure techniques or dedicated sensors.

\myparagraph{Multi-shot}
A typical sensor can capture a wide range of luminances, just not within one shot.
Alternatively, an \emph{exposure sequence}, \ie time-sequential capture of one scene at different exposure settings, can be merged into one image \cite{mann1995being,mitsunaga1999radiometric,debevec1997recovering,robertson2003estimation,granados2010optimal}.
Further, exposure sequences can be fused into a high-quality LDR image \cite{mertens2007exposure,prabhakar2017deepfuse,mustaniemi2020lsd$_2$}.
When dealing with video \cite{kang2003high,kalantari2013patch,gryaditskaya2015motion} or when using neural networks \cite{kalantari2017deep,kalantari2019deep,wu2018deep}, alignment becomes a challenge.

\myparagraph{Single-shot}
Capturing exposure sequences takes time and their alignment is challenging, in particular for video.
This can be alleviated by single-shot solutions relying on custom optics and sensors.
Logarithmic response does not require any exposure control \cite{seger1999hdrc}, but remains prone to noise in dark regions.
Spatially-varying exposure (SVE) techniques place a fixed \cite{nayar2000high,schoeberl2012high,schoeberl2012building,serrano2016convolutional,aguerrebere2014single} or adaptive \cite{nayar2003adaptive,nayar2004programmable} mask of variable optical density in front of the sensor, but face problems with resolution and aliasing.
Beam splitting preserves resolution with different exposures \cite{tocci2011versatile,aggarwal2001split,kronander2013unified} but requires involved optics.
Dual-ISO sensors, \eg Gpixel GMAX and some of the Canon EON sensors, enable varying analogue signal gain for odd and even scanlines.
Their key advantage is that variable blur between scanlines is avoided, as the exposition is fixed for the whole sensor.
On the other hand, instead of collecting more photons in the long exposure and reducing noise this way, only a noisy short exposure is taken, and the long exposure is emulated by increasing ISO, which leads to further noise amplification.
Therefore, denoising and deinterlacing are the key challenges for processing dual-ISO frames \cite{hajisharif2014hdr,go2019image}, including data-driven solutions such as  learned artifact dictionaries \cite{choi2017reconstructing} and CNNs \cite{cogalan2020deep}.
Dual-gain sensors in high-end Canon and professional cinematographic Alexa (ARRI) cameras employ a similar idea but generate two full frames with different analog gains to improve the ratio of read noise to the signal in the high-gain image.
The problem of noise inherent to short exposures, needed to avoid highlight clipping, is reduced by large photosites.

Dual-exposure CMOS sensors
enable varying exposures for odd and even scanlines (some Aptina AR and Sony IMX sensors \cite{sony2020}) or  columns (CMOSIS CMV12000 \cite{cmosis2020}).
\citet{gu2010coded} perform flow-compensated interpolation  for subimage deinterlacing so that differently exposed, full-resolution images are obtained.
\citet{cho2014single} directly calibrate scanlines using bilateral filters followed by motion blur removal \cite{lenzen2011partial} and sharpening. 
Along similar lines, \citet{heide2014flexisp} propose an end-to-end optimization, which jointly accounts for demosaicking, deinterlacing, denoising, and deconvolution.
\citet{an2017single} restore under- and over-exposed pixels using a CNN, but no results for real sensor data are demonstrated.
Our work performs joint denoising, deinterlacing and deblurring, trained on a small set of captured data, resulting in high-quality HDR.

\myparagraph{Exposure on modern sensors}
To better understand the trade-off between single- and dual-exposure sensors, we first conducted a pilot experiment to evaluate the exposure-dependent noise for three different kinds of sensors: 
iPhone (Apple iPhone 8), 
Canon (Canon EOS 550D) and 
Axiom-beta (CMOSIS CMV12000; a full-frame single-exposure setup). For each sensor 600 images of the same scene has been captured in low- and high-exposure (four time longer) modes. An iPhone records 14, Canon and Axiom-beta 12 bits.
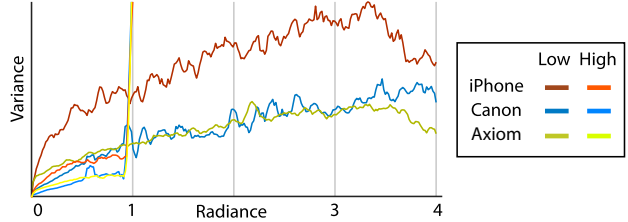
\myfigure{SensorExperiment}{Noise for contemporary sensors at different exposures and intensity:
The horizontal axis is unit radiance.
The vertical axis is variance (less is better).
Different hues depict different sensors.
Bright colors are high, dark colors are low exposure.}
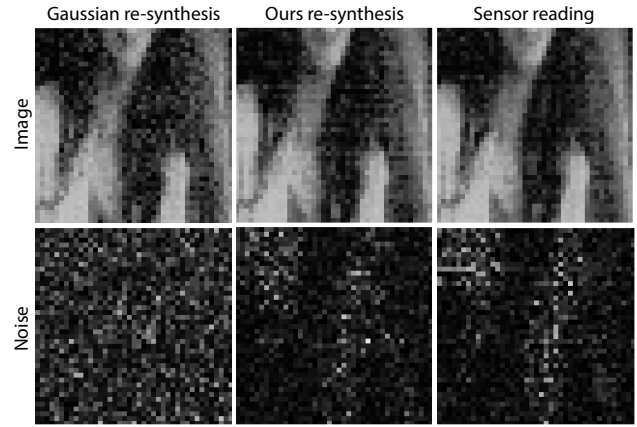
\myfigure{Noise}{A Gaussian noise model \textbf{(left)}, our low-exposure re-synthesis \textbf{(middle)} from a noise-free high-exposure reference (not shown), and a real low-exposure sensor reading reference \textbf{(right).}
Note the long-range correlation across ours and the reference.}
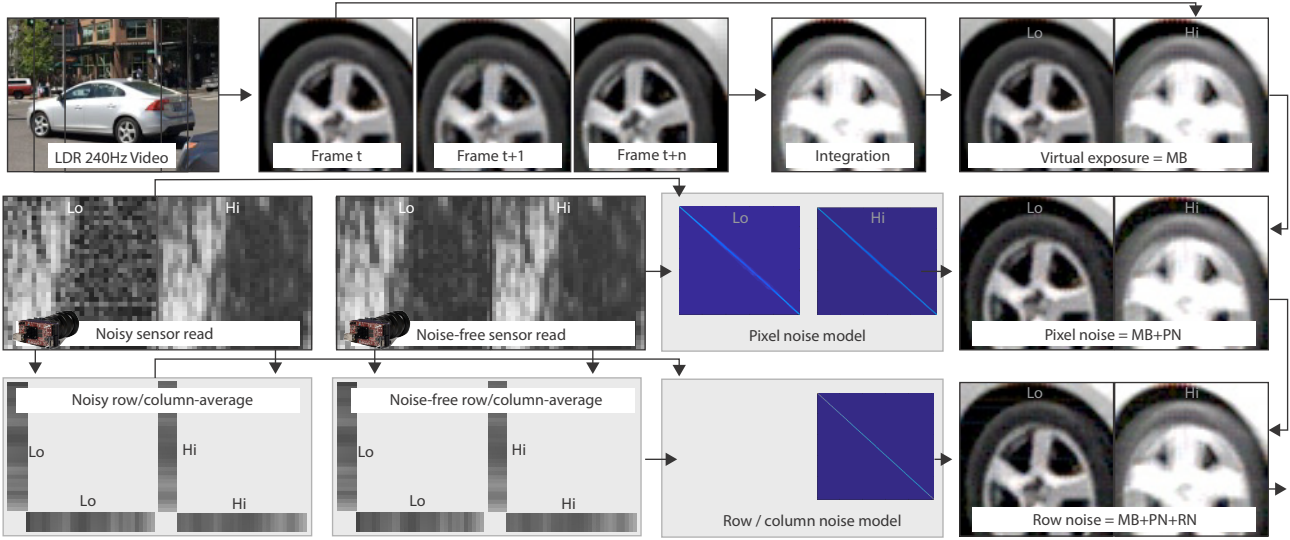
\mycfigure{Pipeline}{Our proposed HDR distortion generation pipeline:
We start from LDR 240~Hz video in the top left, from which frames $t$ to $t+n$ are extracted, integrated, and virtually exposed to produce an image with MB \textbf{(first row)}.
Next, we take pairs of noisy and time-averaged noise-free sensor readings, and produce a non-parametric noise mode (histogram) for low and high exposure.
This noise model is added to the virtual exposure image MB \textbf{(second row)}.
Finally, a model of row and column noise is extracted by averaging vertically or horizontally; this can be added to the pixel noise image, producing the final image with all distortions present \textbf{(third row)}.}
All readings were converted to floating point values between 0 and 4.
High exposure was divided by four to match the same range.
For low exposure, an ideal (as the scene is static) burst fusion was simulated by averaging random four-tuples.
The average of all low-exposure frames is considered the reference for each sensor.
Note, that by construction the reference of the high and low mode is the same.
Then, for every quantized (12 or 14 bit) value $L$ of the reference of each sensor, we select one pixel with that value and compute the variance $\mathrm{Var}(L)$ of all readings in all images.
A high value means that particular sensor for this mode and this absolute radiance has more noise (worse).

\refFig{SensorExperiment} shows that for all sensors, as expected, noise increases with signal \cite{granados2010optimal,janesick2001scientific}.
We further see, that around 0.25 the variance for high exposure diverges (clipping), indicating that these or even higher values cannot be used with long exposure.
More importantly, we also see that low exposure has a higher variance until the point where the high exposure clips.
This trend is true for all sensors, so between 0 and 1: every sensor (hue) at its low exposure mode (brightness) has a higher variance than the high exposure.
This can be attributed to read noise of each burst frame that is accumulated \cite{ma2020quanta}.
This indicates that combining low exposures, even under the ideal condition of no motion, is no immediate solution.
In summary, no single strategy of either averaging low exposures or just using one high exposure is successful across the entire HDR range.
We conclude, that there is a benefit of sensors, which have access to different exposure at different spatial locations.

\mysection{HDR exposure distortion and back}{OurApproach}
Our approach has two steps: learning a model to synthesize distortions to train on (\refSec{CleanToDistorted}; an example result in \refFig{Noise}) and learning to remove distortions (\refSec{DistortedToClean}).

\mysubsection{Clean-to-distorted}{CleanToDistorted}
There are three distortion steps we describe in the order of the underlying physics (\refFig{Pipeline}):
motion blur (\refSec{MotionBlur}),
pixel noise (\refSec{PixelNoise}), and row/column noise (\refSec{RowNoise}).
For all steps, we will look at the analysis from noisy sensor readings to devise a statistical model for inference from \distorted, and a synthesis step to apply it to \clean.

\mysubsubsection{Motion blur}{MotionBlur}
With different exposures in different columns, their MB is also different.
For example, at exposure ratio $r=4$, MB also is four times longer and the image is a mix of sharp and blurry columns.
As getting reference data without MB, in particular HDR, is difficult, we turn to  existing LDR high-speed video footage to simulate multi-exposure MB.

\myparagraph{Data}
We use 123 videos from the Adobe High-speed Video Dataset \cite{su2017deep} which have no, or negligible, inherent MB in a total of 8000 frames.
Note that these are not captured with our sensor, and are LDR.
They are neither input to nor output from of our approach and only provide supervision.

\myparagraph{Synthesis}
Synthesis starts from a random frame of 8-bit LDR high-speed video $I_\mathrm{LDR}$.
It is converted to floating point, and an inverse gamma is applied at $\gamma=2.2$.
We call this the \emph{low frame} image, denoted $I_\mathrm{L} = I_\mathrm{LDR}^\gamma$.
Since our sensor assures that the low and high exposures are ending at the same time \cite{cmosis2020}, to simulate the high frame exposure we average four subsequent $I_\mathrm{L}$, then scale by the exposure ratio, and clamp as in $$
I_\mathrm{H} = \mathtt{clamp}(r\times\mathbb E_{t\in\{0,1,2,3\}} [ I_\mathrm{L}(t)]).
$$
Finally, the low-frame pixels are inserted into the even columns and the high frames into the odd ones, resulting in the motion-blurred image $I_\mathrm{MB}$.

\mysubsubsection{Pixel noise}{PixelNoise}
Pixel noise, which occurs in the sensor, is applied after motion blur, which happens in the optics.
Instead of employing a parametric noise model that has the strengths as priors and for analysis, we use non-parametric histograms to capture a noise model well-suited for generation.
Prior to the noise model derivation, we remove the fixed pattern noise \cite{janesick2001scientific}.

\myparagraph{Data}
We assume we have a limited amount of GT sensor readings available.
In practice, we use no more than 30 pairs of images (not video) captured with the target sensor of everyday scenes, as well as a ground truth acquired by averaging the result of 100 captures of the scene at a very low exposure (so as to make clipping effects negligible) and using a very long exposure.

\myparagraph{Analysis}
The noise is different for different exposures and also for different color channels.
We build a model $p_{c,e}(x|y)$, the probability that when the GT value is $y$, the sensor will read $x$ for channel $c$ and exposure $e$.
A separate model is maintained for every channel in every exposure, leading to six models for three color channels and two exposures.
While we notice the noise models to be similar for different channels at the same exposure, it is, unsurprisingly, different for different exposure.
Histograms $H_{c,e}[x][y]$ are used to represent the probability distribution over $x$ for each $y$ in channel $c$ at exposure $e$.
To construct all histograms, every pair of sensor readings and its ground truth, as well as every pixel and every channel, are iterated, and bin $x$ for histogram $y$ is incremented when the GT pixel is $y$ and the sensor reading is $x$ for channel $c$ and exposure $e$.
The number of histogram bins depends on the bit depth, typically 12 bits, resulting in 4096 bins.
After analysis, all histograms are converted into inverse cumulative histograms $C_{c,e}[x][y]$, allowing us to sample from them in constant time.

\myparagraph{Synthesis}
Noise synthesis is applied to $I_\mathrm{MB}$, the image with simulated MB.
Every pixel and every channel of the MB image $I_\mathrm{MB}$ is iterated to obtain a GT value $y$.
A random number $\xi_{c,e}$ is used to look up  the respective cumulative histogram $C_{c,e}$ to produce a simulated sensor value $x$.
Combining all pixels, channels and exposures results in a virtual synthetic image $I_\mathrm{PN}$ involving MB and pixel noise.

\mysubsubsection{Row/column noise}{RowNoise}
At short exposures more structured forms of noise can become important, one of them being \emph{row/column} noise.
This is not to be mistaken with fixed-pattern noise that frequently is spatially-correlated, but much easier to correct.
In row/column noise, pixels do not change independently; rather, all pixels in a row/column change in correlation, \ie the entire row/column is darkened or brightened.
This is because in the CMOSIS CMV12000 (global shutter) sensor pixel read-out is performed sequentially row-by-row, resulting in differences between the rows.
The analog pixel values are then passed to a column gain amplifier and a column analog-digital converter (ADC), which are used to speed up processing, but introduce differences between the columns \cite{cmosis2020}.
As those effects are visually distracting, we synthesize and ultimately remove them.

\myparagraph{Analysis}
We again iterate all pairs of GT and sensor images, but instead of working on pixels, we now work on entire rows/columns.
In particular we look at the six separate means across every row/column for every channel and exposure.
We denote this mean as $\bar x$ in the sensor image and as $\bar y$ in the GT image.
We now proceed as with pixel noise and build a model in the form of a histogram, resulting in the inverse cumulative row/column noise model $\bar C_{c,e}[\bar x][\bar y]$.

\myparagraph{Synthesis}
Synthesis of row/column noise starts from the image with synthetic MB and pixel noise $I_{PN}$.
We iterate every row, channel and exposure, compute the row/column mean $\bar y_{c,e}$ and again use a random number $\bar\xi_{c,e}$ to draw from $\bar C_{c,e}[\bar\xi][\bar y]$.
To make the row/column mean match the desired mean, we add the difference of the means to the row/column, resulting in the final synthetic noisy image $I_\mathrm{All}$.

\mysubsection{Distorted-to-clean}{DistortedToClean}
We use a U-Net \cite{ronneberger2015u} with residual connections \cite{he2016deep} and sub-pixel convolutions \cite{shi2016real} to map distorted $128\times64\times8$ patches to $128\times128\times 8$ clean patches under an SSIM loss \cite{zhao2016loss} in linear space. Output is converted to RGB and gamma-corrected after the loss.
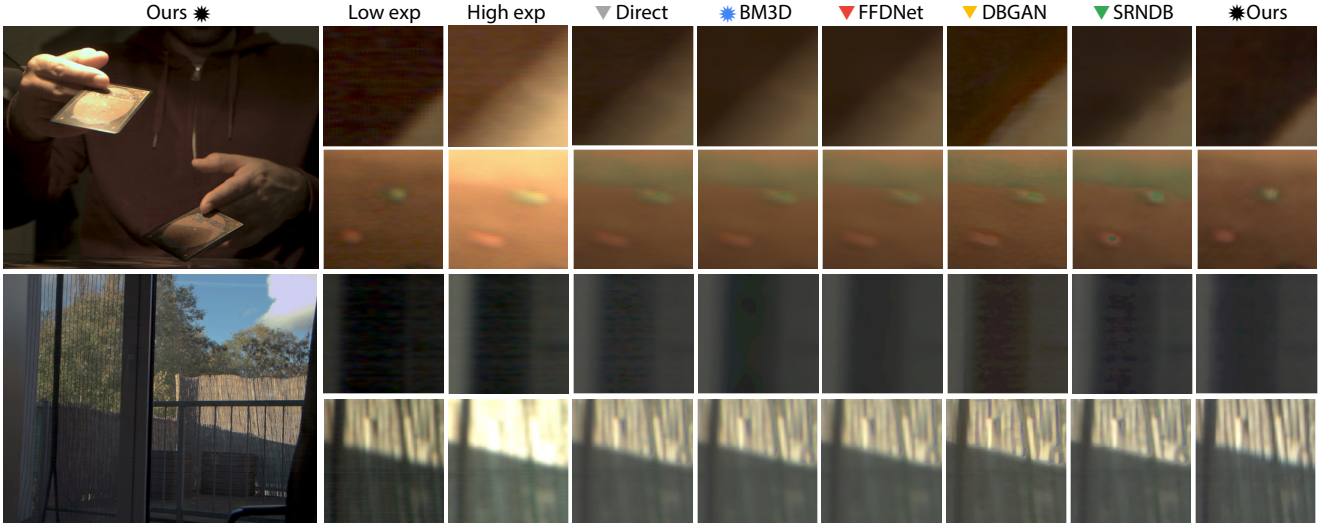
\mycfigure{Results}{Comparison of different methods \textbf{(columns)} on two scenes \textbf{(rows)}.
Please see the text for discussion.}

\mysection{Results}{Results}
We present quantitative and qualitative evaluation on deblurring/denoising (\refSec{Quantitative}), super-resolution (\refSec{TemporalResults}), and HDR illumination reconstruction (\refSec{Envmap}) tasks. Interactive comparison and videos can be found at \url{https://deephdr.mpi-inf.mpg.de}.

All test images have been captured using an Axiom-beta camera with a CMOSIS CMV12000 sensor \cite{cmosis2020} and a Canon EFS 18-135\,mm lens at resolution 4096 $\times$ 3072 RAW 12-bit pixels, using the lowest gain with exposure ratio 4 (or 16 when explicitly mentioned) and (low) exposure time varying from 1 to ~8\,ms.
Although our noise model is created for a given fixed ratio, the exposure times for the two discrete exposures can vary continuously as we show in the supplemental materials.
All results are shown after gamma correction and photographic tone mapping \cite{reinhard2002photographic}. CMOSIS CMV12000 sensor \cite{cmosis2020} is a CMOS sensor that features global shutter, large pixel sizes, low dark current noise, and is relatively inexpensive in comparison with CCD sensors with similar performance.
Therefore the sensor is suitable for demanding computer vision applications and it is offered by many well-known industrial camera makers  \cite{basler2020, omnivision2020, emergentvision2020}. 

\mysubsection{Denoising/deblur evaluation}{Quantitative}
We now evaluate the combination of our method and our synthetic training data as well as other ways to obtain training data and other methods for denoising and deblurring.

\myparagraph{Methods}
We consider eight methods (color-coded; ``Method'' in \refTbl{Results}):
\nameInColor{Direct} is a non-learned direct, physics-based fusion of the low and high frame, with bicubic upsampling \cite{debevec1997recovering}.
Next, \nameInColor{BM3D}~\cite{dabov2007image} is a gold-standard, non-deep denoiser.
When BM3D is ``trained'' this means performing a grid search on the training data in order to find the standard deviation parameter with the the optimal DSSIM.
\nameInColor{FFDNet}~\cite{zhang2018ffdnet} is a state-of-the-art deep denoiser.
\nameInColor{DBGAN}~\cite{kupyn2019deblurgan} and \nameInColor{SRNDB}~\cite{tao2018scale} are recent deblurring approaches.
\nameInColor{LSD}~\cite{mustaniemi2020lsd$_2$} is a deep multi-exposure method that produces denoised and deblurred LDR images. 
The final method is \nameInColor{Heide}~\cite{heide2014flexisp} which is a general image reconstruction method, capable of working with multiplexed exposures.

\myparagraph{Training data}
For each  method, we study how it performs when trained with different data (``Train. data'' in \refTbl{Results}).
Each type of training data has a different symbol.
We denote it as \nameAndIcon{Theirs} if the authors provide a pre-trained version.
\nameAndIcon{Sensor} means training on the image for which we have paired training data available directly, \ie without our proposed re-synthesis.
Please note that this training is not applicable to tasks that involve removing MB, as the supervision inevitably contains MB.
Next, we study heteroscedastic Gaussian noise, \nameAndIcon{HetGau} which refers to taking our training data, fitting a linear model of Gaussian parameters of the error distribution  and then re-synthesizing training.
Finally, we study four ablations of our training data generation:
only motion blur \nameAndIconA{OurMB}, only pixel noise \nameAndIconA{OurPN}, only row noise \nameAndIconA{OurRN}, and finally \nameAndIconA{OurAll} in \refTbl{Results}.

\myparagraph{Metrics}
We measure DSSIM \cite{wang2004image}, where less is better.

\myparagraph{Tasks}
We study four tasks (four last columns in \refTbl{Results}):
First, we remove noise in the low exposure only (\task{Lo2Lo}).
Second, we remove noise and MB in the high exposure only (\task{Hi2Hi-MB}).
Third, is a task where input is both exposures and output is an HDR image without noise, \task{LoHi2HDR}.
The fourth task consumes low and high exposures, and removes both noise and MB to output HDR (\task{LoHi2HDR-MB}).
In all tasks, the exposure ratio, 1:4, is in favor of competitors conceived for LDR use.
The test set for all tasks contains 10 images.

\begin{table}[]
    \centering
    \setlength{\tabcolsep}{4pt}
    \caption{Performance of different methods and different training data \emph{(rows)} for different tasks \emph{(columns)}.
    Different icon shapes denote different training; colors map to different methods.}
    \vspace{0.1cm}
    \begin{tabular}{lc rrrr}
        \toprule
        &
        &
        \multicolumn4c{Task}
        \\
        \cmidrule(lr){3-6}
        &%
        \multicolumn1r{In Lo}&
        \multicolumn1c\cmark&
        \multicolumn1c\xmark&
        \multicolumn1c\cmark&
        \multicolumn1c\cmark\\
        &
        \multicolumn1r{In Hi+MB}&
        \multicolumn1c\xmark&
        \multicolumn1c\cmark&
        \multicolumn1c\cmark&
        \multicolumn1c\cmark\\
        &
        \multicolumn1r{Out MB}&
        \multicolumn1c\xmark&
        \multicolumn1c\xmark&
        \multicolumn1c\cmark&	
        \multicolumn1c\xmark\\
        &
        \multicolumn1r{Out HDR}&
        \multicolumn1c\xmark&
        \multicolumn1c\xmark&
        \multicolumn1c\cmark&
        \multicolumn1c\cmark\\
        \toprule
        \multicolumn1c{Train. data}&
        \multicolumn1c{Method}&
        \multicolumn4c{Error (DSSIM$\times10^{-2}$)}\\
        \toprule
\colorIconAndName{Direct}{Theirs} & \multirow{1}{*}{Direct \cite{debevec1997recovering}}& 7.87&
7.08&
3.70&
5.52
\\
\midrule
\colorIconAndName{BM3D}{Theirs}&
\methodCell{5}{BM3D \cite{dabov2007image}}& 
2.98&
4.10&
2.00&
2.63
\\
\colorIconAndName{BM3D}{Sensor}&&
2.84&
\nacell&
1.90&
\nacell\\
\colorIconAndName{BM3D}{HetGau}&&
2.75&
3.86&
1.76&
2.32\\
\colorIconAndName{BM3D}{OurAll}&&
\bf{2.72}&
3.93&
1.80&
2.35
\\
\midrule
\colorIconAndName{FFDNet}{Theirs}&
\methodCell{3}{FFDNet \cite{zhang2018ffdnet}}&
3.79&
4.31&
2.18&
2.83
\\
\colorIconAndName{FFDNet}{Sensor}&&
2.78&
\nacell&
2.03&
\nacell\\
\colorIconAndName{FFDNet}{OurAll}&&
2.78&
3.92&
2.03&
2.54
\\
\midrule
\colorIconAndName{DBGAN}{Theirs}&
\methodCell{1}{DBGAN \cite{kupyn2019deblurgan}}&
5.31 & 4.88 & 2.95 & 3.32 \\
\midrule
\colorIconAndName{SRNDB}{Theirs}&
\methodCell{1}{SRN-DB \cite{tao2018scale}}&
3.28 & 4.36 & 2.27 & 2.60 \\
\midrule
\colorIconAndName{LSD}{Theirs}&
\methodCell{1}{LSD$_2$ \cite{mustaniemi2020lsd$_2$}}&
\nacell & 2.94 & 3.24 & 2.46 \\
\midrule
\colorIconAndName{Heide}{Theirs}&
\methodCell{1}{Heide et al. \cite{heide2014flexisp}} &
\multicolumn{4}{c}{5.27} \\
\midrule
\colorIconAndName{Our}{Sensor}&
\methodCell{6}{Ours}&
6.51&
\nacell&
4.62&
\nacell
\\
\colorIconAndName{Our}{HetGau}&&
3.14& 3.17& 2.35 &2.15\\
\colorIconAndName{Our}{OurRN}& & 5.33 & 5.24 & 4.41 & 4.32 \\
\colorIconAndName{Our}{OurPN}& & 4.24 & 4.60 & 3.01 & 3.06 \\
\colorIconAndName{Our}{OurMB}& & 4.23 & 3.63 & 2.17 & 3.15 \\
\colorIconAndName{Our}{OurAll} & & 2.75 & \bf{2.64} & \bf{1.68} & \bf{1.84} \\
        \bottomrule
    \end{tabular}
    \label{tbl:Results}
\end{table}

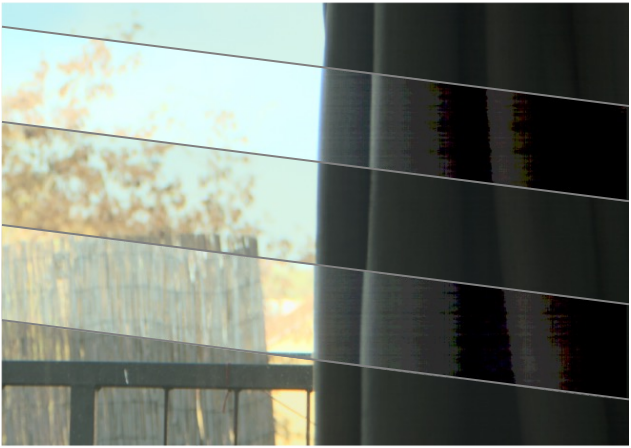
\myfigure{16x}{Comparison of our reconstruction at an exposure rate of 16:1 and the best single exposure result \textbf{(inset stripes)}.}
\myparagraph{Discussion}
Results are shown in \refTbl{Results}.
Our method trained on our synthetic training data (\colorIcon{Our}{OurAll}) performs best on all tasks.
Our ablations (\colorIcon{Our}{OurRN}, \colorIcon{Our}{OurPN} and \colorIcon{Our}{OurMB}) all perform worse than the full method, indicating all additions are relevant.
Looking into how other methods trained on data synthesized using our distortion model perform (\colorIcon{BM3D}{OurAll} and \colorIcon{FFDNet}{OurAll}), we see that first, they all improve in comparison to being trained on their original data (\colorIcon{BM3D}{Theirs}, \colorIcon{FFDNet}{Theirs}, \colorIcon{DBGAN}{Theirs}, \colorIcon{SRNDB}{Theirs} and \colorIcon{LSD}{Theirs}, respectively), but, second, none can compete with our method trained on that data (\colorIcon{Our}{OurAll}).
Only \colorIcon{BM3D}{OurAll}, as a competing method, when tuned on our data, can compete on its home ground, \task{Lo2Lo}.
We also tried training our network with other data, such as using
sensor data directly (\colorIcon{Our}{Sensor}), 
hetroscedatic Gaussian noise (\colorIcon{Our}{HetGau}), but none of these was able to capture the combination of motion blur, pixel noise and row/column noise, resulting in larger errors.
As a sanity check, we also tuned BM3D on 
sensor data (\colorIcon{BM3D}{Sensor}) and
hetroscedatic Gaussian noise (\colorIcon{BM3D}{HetGau}), but no choice of parameters, even with that information, can get BM3D to perform much better on test data.
A further test is to compare to \colorIcon{Direct}{Theirs}, which is not learned or doing anything except up-sampling and fusion; this should be a lower bound for any method or task.
Finally, our approach compares favorably to \citet{heide2014flexisp} (\colorIcon{Heide}{Theirs}), a general, powerful and flexible imaging framework that can work on multi-exposure images.
When looking at performance for different tasks, we find that for simpler tasks, such as \task{Lo2Lo}, \ie a direct denoising, unsurprisingly, our best result (\colorIcon{Our}{OurAll}) performs comparably to the gold standard (\colorIcon{BM3D}{Theirs}), in particular when tuned on our data (\colorIcon{BM3D}{OurAll}).
When the task gets more involved, \ie removing MB or producing HDR, the methods start to perform more similarly, but ours tends to win by a larger margin.
For completeness, our analysis includes methods designed for denoising being applied to a deblurring task or vice versa.
As all tasks except \task{Lo2Lo} involve components of both deblurring and denoising, we report those numbers to certify that no method solving only one of the tasks, does it so well that the DSSIM is reduced more than another method trying to solve both tasks.
This is probably because both noise and blur are visually important, and no method, including ours, can reduce one of them enough to make the other irrelevant.
In summary, using the right training data helps our methods and others to solve multiple aspects of multiple tasks.

The quantitative results from above are complemented by the qualitative ones in \refFig{Results}.
The first row shows our (\colorIcon{Our}{OurAll}) complete image.
The second and third row show selected patches from the the low and high input, which suffer from noise or blur respectively.
Directly (\colorIcon{Direct}{Theirs}) fusing both into HDR, as in the fourth column, reduces noise and blur, but cannot remove them.
The BM3D (\colorIcon{BM3D}{OurAll}) and FFDNet (\colorIcon{FFDNet}{Theirs}) columns show that individual frames can be denoised, but blur remains.
This is most visible in moving parts, such as the dots in the second row.
Using de-blurring, as in DBGAN (\colorIcon{DBGAN}{Theirs}) or SRNDB (\colorIcon{SRNDB}{Theirs}), can reduce blur, but this often leads to ringing.
Our joint method (\colorIcon{Our}{OurAll}) performs best on these.

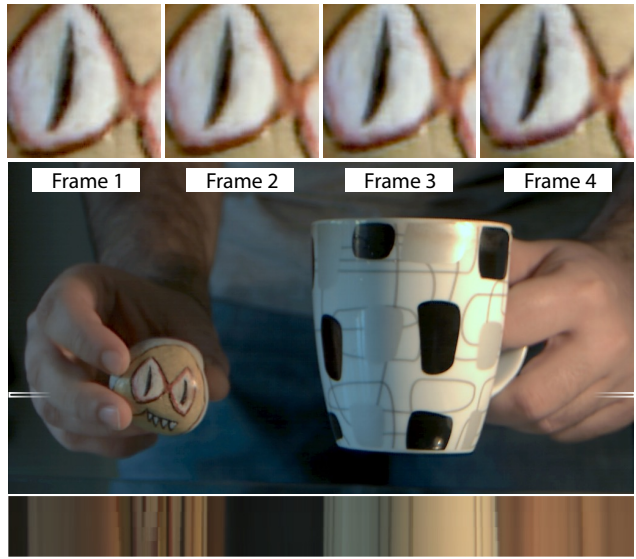
\myfigure{Temporal}{ Four frames cropped \textbf{(top)} from an HDR video with temporal super-resolution using \method{Our} approach.
The full frame 2 \textbf{(middle)}.
An epipolar slice for the marked row \textbf{(bottom)}.}
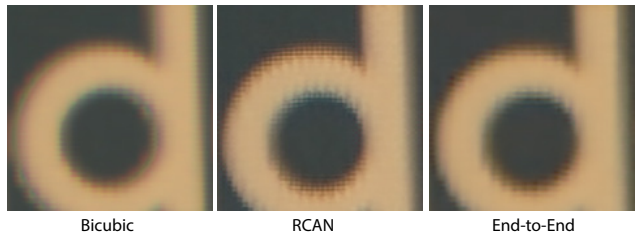
\myfigure{Superresolution}{Spatial super-resolution.}
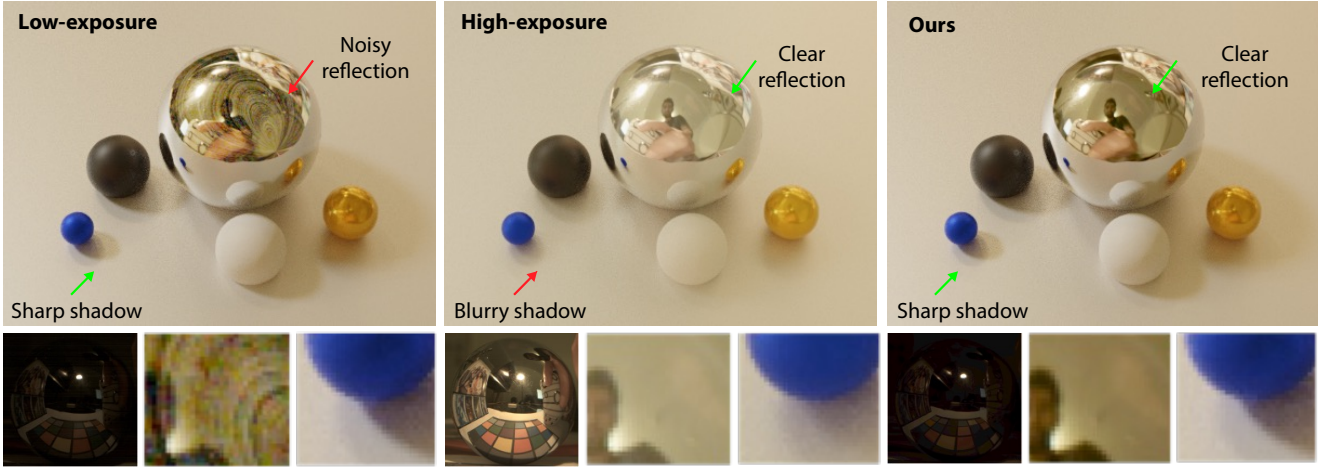
\mycfigure{Rendering}{Rendering from a spherical illumination map captured at a low exposure \textbf{(left)}, a high exposure \textbf{(middle)} and using our approach \textbf{(right)}.
For each approach the illumination is seen as an inset on the left.
For the low exposure, the shadows are sharp, as the light source did not saturate, but the dark regions are clipped and massively noisy.
For the high exposure, the dark regions are reproduced, slightly noisy, but the light source is clamped, leading to a loss in dynamic range and a loss of sharp shadows.
Our method reproduces both.
Note that visible overall brightness differences are expected, as clamping is present in some images, which does not conserve energy.}
\refFig{16x} compares our result at an exposure rate of 16:1 to the best single-exposure result.
We note our approach reproduces details in the bright (outdoor) part as well as in the dark (indoor) part despite the massive contrast. 
The best LDR fit can resolve some of the outdoor elements, but has no details except quantization noise in the dark part.

\mysubsection{Temporal and spatial super-resolution}{TemporalResults}
In temporal super-resolution \cite{jiang2018super}, we extend the \task{LoHi2Hdr-Mb} task to output not a single image, but $n$ images instead.
To generate training data, we still extract sequences of $n$ high-speed video frames, and we still call the first frame the low frame and the integral of all $n$ frames the high exposure. 
The architecture is identical, except that it produces $n$ images in the last layer.
Note that the input is still only two interleaving exposures, where one has severe MB and the other severe noise. \refFig{Temporal} shows \method{Our} End-to-End reconstruction, and in the supplemental materials we demonstrate a continuous adjustment of blur magnitude.


\begin{table}[]
    \centering
    \caption{
    HDR super-resolution in combination with denoising and deblurring.
    ``Us-Them'' means to first run \method{Our} non-super-resolution method, followed by: temporal  \cite{jiang2018super}, or spatial \cite{zhang18} super-resolution method.
    ``End-to-End'' means \method{Our} full method.
    }
    \vspace{.2cm}
    \label{tbl:Temporal}
    \begin{tabular}{lcc}
        \toprule
        &
        Us-Them&
        End-to-End\\
        \midrule
        Temporal&
        0.032&
        0.026\\
        Spatial&
        0.043&
        0.035\\
        \bottomrule
    \end{tabular}
\end{table}

Analogously to temporal super-resolution, we can also look at spatial super-resolution \cite{zhang18}.
Here, training data is spatially down-scaled before being used to simulate Hi and Lo frames.
At training time, the decoder branch is simply repeated several times to produce output patches larger than the input patches.
\refFig{Superresolution} compares a bicubic upsampling and RCAN \cite{zhang18} to \method{Our} End-to-End method.

\refTbl{Temporal} compares our methods with  \cite{jiang2018super} and  \cite{zhang18}.

\mysubsection{Application: HDR illumination reconstruction}{Envmap}
A key application of HDR  is to use it for illumination  \cite{debevec1997recovering}.
We captured a mirror ball, removed motion blur and noise using  our full method (\colorIcon{Our}{OurAll}), and re-rendered it using Blender's \cite{blender2020blender} path tracer with 512 samples and automatic tone and gamma mapping.
The resulting image is seen in \refFig{Rendering}.
We find that the non-linear mapping of MC rendering amplifies structures and noise gets more visible, in particular row noise.
Using only the high exposure removes noise, but cannot capture the dynamic range, resulting in washed-out shadows.
Our method succeeds in removing it, in particular row noise, resulting in sharp shadows as well as noise-free reflections.
Note that some noise is present in all images due to finite MC sample count (all images computed 20 min.).
The noise appears less in the high exposure, as reduced contrast results in an easier light simulation problem that leads to an overall incorrect, strongly biased solution.

\mysection{Conclusion}{Conclusion}

We presented a CNN solution for HDR image reconstruction tailored for a single-shot dual-exposure sensor.
By joint processing of low and high exposures and taking advantage of their perfect spatial and temporal registration, our solution solves a number of serious problems inherent to such sensors such as correlated noise and spatially varying blur, as well as interlacing and spatial resolution reduction.
We demonstrate that, by capturing a limited amount of data specific for such sensors and using simple histograms to represent the noise statistics, we were able to generate synthetic training data that led to a better denoising and deblurring quality than achieved by existing state-of-the-art techniques.
Moreover, we show that by using our limited sensor-specific data, the performance of other techniques can greatly be improved.
This is for two reasons: First, previous methods did not have access to massive amounts of training data for dual-exposure sensors, a problem we solve here by proposing the first dedicated distortion model allowing to synthesize training data.
Second, dual-exposure sensors in combination with proper CNN-based denoising and deblurring provide us with much richer data managed to fuse.

\clearpage
{\small
\bibliographystyle{ieee_fullname}
\bibliography{egbib}
}

\end{document}